\documentclass[prb,aps,twocolumn,showpacs,preprintnumbers,amsmath,amssymb,10pt]{revtex4-1}
\usepackage{graphicx}
\usepackage{dcolumn}
\usepackage{bm}
\usepackage{ulem}
\begin{document}


\title{Andreev-Lifshitz Supersolid Hydrodynamics Including the Diffusive Mode} 
\author{ Matthew R. Sears } 
\author{ Wayne M. Saslow } 
\email{wsaslow@tamu.edu}
\affiliation{ Department of Physics, Texas A\&M University, College Station, TX 77843-4242 }
\date{\today}

\begin{abstract}
We have re-examined the Andreev-Lifshitz theory of supersolids.  This theory implicitly neglects uniform bulk processes that change the vacancy number, and assumes an internal pressure $P$ in addition to lattice stress $\lambda_{ik}$.  Each of $P$ and $\lambda_{ik}$ takes up a part of an external, or applied, pressure $P_a$ (necessary for solid $^4$He).  The theory gives four pairs of propagating elastic modes, of which one corresponds to a fourth-sound mode, and a single diffusive mode, which has not been analyzed previously.  The diffusive mode has three distinct velocities, with the superfluid velocity much larger than the normal fluid velocity, which in turn is much larger than the lattice velocity.  The mode structure depends on the relative values of certain kinetic coefficients and thermodynamic derivatives.  We consider pressurization experiments in solid $^4$He at low temperatures in light of this diffusion mode and a previous analysis of modes in a normal solid with no superfluid component.  
\end{abstract}



\maketitle
\section{Introduction}

Since the late 1960's there have been theoretical suggestions that solids might display flow behavior similar to what is found in superfluids.\cite{AL69,Thouless69,Chester70,Leggett70}  For that reason there has been a great deal of interest in solid $^4$He as a candidate to be a {\it supersolid}.\cite{BalCau08}  The first experimental indication of superflow was the appearance of a non-classical moment of inertia (NCRI), first observed by Chan's group, since confirmed by many other laboratories, and strongly linked to disorder.\cite{KC1,KC2,RR1,Shira1,Koj1,Penzev,RR2,RR3,Lin07,ClarkWestChan07}  In addition, the shear modulus shows anomalous behavior,\cite{DayBeamishNature07} although not enough to explain the NCRI experiments.\cite{ChanScience08}  Non-NCRI superflow has  been searched for but not observed.\cite{Day06}  Evidence is growing that restricts the possible temperature range over which supersolidity can occur.\cite{ShevDayBeam10}  Moreover, for NCRI experiments with rim velocity $v$ at temperature $T$, the observed hysteresis in $v-T$ space suggests multiple apparent phase transitions.\cite{Koj1,Hunt09,Choi10}

A recent experiment on a pancake-shaped sample, where a pressure change is applied to one side, finds an exponential decay with time of the pressure response on the opposite side.\cite{RitRep09}  The response is slower at lower temperatures, rather than saturating as for a quantum transition, perhaps an indication that the system is {\it not} supersolid.  We have recently studied the lattice diffusion mode of a normal solid (see Ref.\onlinecite{SearsSasALNS10}), based on equations obtained by eliminating the superfluid velocity $\vec{v}_{s}$ from the theory of Andreev and Lifshitz.\cite{AL69}  We obtained both the diffusion constant and the eigenmode structure (by which we mean the ratios of the deviations from equilibrium of various thermodynamic quantities) for a solid under an externally applied pressure $P_a$ (necessary to solidify $^4$He, even at $T=0$).  



Whereas an ordinary solid has eight degrees of freedom,\cite{SearsSasALNS10} the addition of $\vec{v}_{s}$ (the gradient of a phase) gives a supersolid nine degrees of freedom.  For a plane wave, where $k_i$ is the wavevector with magnitude $k$) the degrees of freedom are given by two scalar thermodynamic quantities (which can be taken to be the mass density $\rho$ and the entropy density $s$), the lattice vector $u_{i}$, the normal fluid velocity vector ${v_n}_{i}$, and ${v_s} = k_i {v_s}_i / k$, where ${v_s}_i$ is the superfluid velocity.  The scalar quantity $v_s$ has been defined because ${v_s}_i$ is expressible as the gradient of a phase $\phi$.  The total momentum density is thus given by $g_{i}=\rho_n {v_n}_{i} + \rho_s {v_s}_i$, where $\rho_n$ and $\rho_s$ are the respective densities of the normal and superfluid components.  (In principle, both $\rho_n$ and $\rho_s$ are tensors, but calculations for hcp $^4$He indicate that they are nearly isotropic.\cite{SasJolad,GalliSas})  The nine degrees of freedom imply there are nine normal modes.  For a uniform infinite system these modes are: four pairs of propagating elastic waves (previously studied for both zero \cite{AL69,Saslow77,Liu,YooDorsey} and nonzero\cite{SearsSasSSGen} $P_a$), with frequency $\omega\sim k$; and a diffusive mode, with $\omega\sim ik^{2}$ (whose structure has not been previously studied).

The present work studies this diffusive mode that occurs in a supersolid when $\vec{v}_{s}$ is included.  We employ a variation on the notation of Ref.\onlinecite{Saslow77}, which gives a more explicit derivation of the equations of motion than does Ref.\onlinecite{AL69}, and extends Ref.\onlinecite{AL69} to include nonlinear terms.\cite{SaslowNote}  Ref.\onlinecite{AL69} and Ref.\onlinecite{Saslow77} implicitly assume that uniform vacancy-number-changing bulk processes are negligible.  

The Andreev-Lifshitz theory is remarkable in that it assigns an internal pressure $P$, in addition to lattice stress $\lambda_{ik}$, to a supersolid, in order to continuously go to the superfluid limit.  Each of $P$ and $\lambda_{ik}$ take up part of $P_a$.  Ref.~\onlinecite{SearsSasALNS10} finds for a solid, by thermodynamic considerations, the dependence of $P$ on $P_a$.  The consequences of distinct $P$, $\lambda_{ik}$, and $P_a$ had not previously been considered.  Ref.~\onlinecite{SearsSasALNS10} calculates the effect of $P_{a}$ on the propagating elastic and diffusive modes of an ordinary solid.  Ref.~\onlinecite{SearsSasSSGen} calculates the effect of $P_{a}$ on the propagating elastic modes of a supersolid, as well as the efficiency with which a heater or a transducer generates these modes. The present work considers the effect of $P_{a}$ on the diffusive mode of a supersolid.


As for the lattice diffusion mode for the normal solid, the diffusive mode for the supersolid is characterized not by the diffusion of a single thermodynamic variable, but by specific amounts of each, determined by the eigenmode structure.  A dissipative term in the equation of motion for the lattice displacement permits the lattice velocity to differ from $\vec{v}_{n}$.  We find the relationships between the normal, superfluid, and lattice velocities in this mode. From the lattice velocity one can obtain the lattice displacement and lattice strain deviation.  Because the mode is diffusive, the rate of change of the momentum density, and thus the total stress deviation, are nearly zero, so lattice stress deviations must be canceled by an opposing pressure deviation $P'$, thus determining $P'$.  Again because the mode is diffusive, the rate of change of $\vec{v}_{s}$, and thus the total chemical potential deviation, are nearly zero, so the $P'$ must be compensated by an opposing contribution due to a temperature deviation $T'$.  This diffusion mode is therefore characterized by its diffusion constant and specific ratios of the normal and superfluid velocities, and the temperature and pressure, relative to the lattice velocity.  In practice we use the entropy and mass densities rather than temperature and pressure.  The theory permits vacancies to diffuse but there are no bulk sources or sinks for them.

Section~\ref{ALtheory} gives the AL supersolid theory in our notation.  Section~\ref{NormModes} derives the normal modes for the supersolid.  Section~\ref{Summary} provides a summary.  Appendix~\ref{PaTOKappendix} estimates the sizes of several quantities relevant to the diffusive mode.

\section{Andreev-Lifshitz Supersolid}
\label{ALtheory}
In what follows we employ the primary quantities energy density $\epsilon$, lattice displacement $u_{i}$, and non-symmetrized strain $w_{ik}=\partial_{i}u_{k}$. 

\subsection{Thermodynamics}
\label{ThermoSubsection}
The thermodynamic equations for a supersolid are given by 
\begin{align}
d\epsilon &= Tds+\lambda_{ik}dw_{ik}+ \mu d\rho +\vec{v}_n\cdot d\vec{g} +
\vec{j}_s \cdot d\vec{v}_s, \label{s-depsilon}\\ 
\epsilon &= -P+Ts+\lambda_{ik}w_{ik}+\mu \rho +\vec{v}_n \cdot\vec{g}+ \vec{j}_s
\cdot \vec{v}_s, \label{s-epsilon}\\
0 &= -dP+sdT+w_{ik}d\lambda_{ik}+\rho d\mu +\vec{g}\cdot d\vec{v}_n + \vec{v}_s
\cdot d\vec{j}_s. \label{s-GibbsDuhem}
\end{align}
Here $\lambda_{ik}$ is an elastic tensor density (with the same units as pressure $P$), $\mu$ is the chemical potential (with units of velocity squared), 
\begin{equation}
 \vec{g} = \rho_n {\vec{v}_n} + \rho_s {\vec{v}_s}
\end{equation}
is the momentum density, and 
\begin{equation}
 \vec{j}_s = \vec{g} - \rho \vec{v}_n = \rho_s (\vec{v}_s - \vec{v}_n).
\end{equation}
is a momentum density defined so that $d \epsilon = \vec{g} \cdot d(\delta \vec{v})$ under a Galilean boost $\delta \vec{v}$.  Since $(\vec{g}, \vec{v}_{n}, \vec{v}_{s})$ are all vectors under Galilean boosts, we deduce that $\rho_n+\rho_s=\rho$.

We find it convenient to define 
\begin{equation}
 \vec{j}_n \equiv \rho \vec{v}_n, \label{jn}\\
\end{equation}
so that 
\begin{equation}
\vec{g} = {\vec{j}}_n +  {\vec{j}}_s.
\label{gjnjs}
\end{equation}
Unlike $\vec{j}_{s}$, the quantity $\vec{j}_{n}$ is a momentum density both in units {\it and} in its properties under Galilean boosts.  

\subsection{Dynamics}
\label{DynamicsSubsection}
The linearized equations of motion relevant to obtaining the normal modes, considering only the independent variables $s$, $u_{i}$, $\rho$, ${v_n}_i$, and ${v_s}_i$, are
\begin{align}
\partial_{t}s+\partial_{i}f_{i}&=\frac{R}{T},\quad (R \geq 0) \label{s1}\\
\partial_{t}u_{i}&=U_{i}, \label{u1}\\
\partial_{t}\rho+\partial_{i}g_{i}&=0, \label{rho1}\\
\partial_{t}g_{i}+\partial_{k}\Pi_{ik}&=0, \label{g1}\\
\partial_{t}{v_s}_i + \partial_i \theta &=0. \label{vs1}
\end{align}
Here, the fluxes $f_{i}$, $\Pi_{ik}$, $g_{i}$, $\theta$, and the ``source'' $U_{i}$ are given by
\begin{align}
f_{i}&=s{v_n}_i-\frac{\kappa_{ij}}{T}\partial_{j}T-\frac{\alpha_{ij}}{T}
\partial_{l}\lambda_{lj}, \label{s-f}\\
U_{i}&={v_n}_{i}+\frac{\alpha_{ij}}{T}\partial_{j}T+\beta_{ij}\partial_{l}
\lambda_{lj}, \label{s-U}\\
\Pi_{ik}&=(P\delta_{ik}-\lambda_{ki})-\eta_{iklm}\partial_{m}{v_n}_{l} -
\zeta_{ik}\partial_{l}{j_s}_l, \label{s-Pi}\\
\theta &= \mu - \zeta_{ik} \partial_k {v_n}_i - \chi \partial_k {j_s}_k,
\label{s-theta}\\
g_{i}&=\rho {v_n}_{i} + {j_s}_i, \label{s-g}
\end{align}
and we take $R \approx 0$, as it is second order in deviations.  AL use both $\sigma_{ik} \approx - \Pi_{ik}$ (a notation we employ below) and 
$j_i = g_i$.  The term in \eqref{s-U} proportional to $\beta_{ij}$ allows the lattice velocity $\dot{u}_{i}$ to differ from the velocity ${v_n}_{i}$ associated with mass flow.  

Recall that a diffusion constant $D$ is proportional to a characteristic velocity times a characteristic mean-free path, so it has units of m$^{2}$/sec.  In terms of a $D$, the dissipative coefficients have the following units: $\kappa_{ij}$ has units of $s$ times $D$; $\alpha_{ij}$ has units of $D$; $\beta_{ij}$ has units of inverse pressure times $D$; $\eta_{iklm}$ has units of $\rho$ times $D$; $\zeta_{ik}$ has units of $D$; and $\chi$ has units of inverse density times $D$. 

\section{Normal Modes in a Supersolid}
\label{NormModes}
As noted earlier, this system has nine variables: $s$, $\rho$, $u_{i}$, ${v_n}_{i}$, and $v_s$.  With deviations from equilibrium denoted by primes, we use the nine variables $s'$,
$\rho'$, $u'_{i}$,
\begin{equation}
g_i' \approx\rho_n {v_n}'_{i} + \rho_s {v_s}'_{i},
\end{equation}
and
\begin{equation}
v_{s}' = \frac{k_i {v_s}'_i}{ k}.
\end{equation}
As noted above, there correspondingly are nine normal modes.  For an infinite system we assume a disturbance of the form $\exp[i(\vec{k}\cdot
\vec{r}-\omega t)]$, where the real wavevector $\vec{k}$ is considered to be known, but $\omega$ is unknown.  For the disturbance to decay in time, $Im(\omega)<0$.  We find that six modes come in three degenerate pairs, with $g_{i}'$ and $u_{i}'$ strongly coupled, and correspond to ordinary elasticity.  Two other modes also form a degenerate pair, corresponding to fourth sound, with the superfluid component in motion and the normal component essentially at rest.\cite{Atkins59,RudnickShapiro62,SearsSasSSGen} The ninth and final mode is diffusive, with $v_n'$ and $v_s'$ in opposing directions, and nearly constant chemical potential and stress. 

We consider the (off-diagonal) temperature-lattice transport coefficient $\alpha_{ij}=0$, and set to zero the thermal expansion coefficient.  We also neglect the viscosities $\eta_{iklm}$, $\zeta_{ik}$, and $\chi$, which  to lowest order do not contribute to the modes.\cite{ViscosityFootnote}  We consider an isotropic solid, for which $\kappa_{ij}=\kappa\delta_{ij}$ and $\beta_{ij}=\beta\delta_{ij}$.  

Unless otherwise specified, thermodynamic derivatives with respect to $\rho$, $s$, or $w_{ik}$ are taken with the other two variables held constant. 

\subsection{Elastic Modes}
\label{ElasticSubsection}
The elastic modes are obtained by neglecting dissipative and nonlinear terms in \eqref{s1}-\eqref{vs1}.  
Although the elastic modes of a supersolid had previously been found for $P_a=0$,\cite{AL69,Saslow77,Liu}  Ref.~\onlinecite{SearsSasSSGen} explicitly finds the elastic modes for nonzero $P_a$ (recall that a $P_a \gtrsim 25$~bars is necessary to solidify $^4$He).  A summary of the results and convenient notation are provided here.  

For the isotropic case, we define
\begin{align}
\frac{\partial P}{\partial w_{ik}} \equiv& \frac{\partial P}{\partial w} \delta_{ik},\label{dPdw}\\
\frac{\partial \lambda_{ik}}{\partial \rho} \equiv& \frac{\partial \lambda}{\partial \rho} \delta_{ik},\label{dlambdadrho}\\
\frac{\partial \lambda}{\partial w} \equiv& K + \frac{4}{3} \mu_V. \label{dlambdadw}
\end{align}
In this case the static value of the strain (dependent on the applied pressure) is isotropic:\cite{SearsSasALNS10,LLElasticity}
\begin{align}
w_{ik}^{(0)} = \frac{w_{ll}^{(0)}}{3} \delta_{ik} \approx -\frac{P_a}{3K} \delta_{ik}.
\label{staticstrain}
\end{align}


\subsubsection{Longitudinal Elastic Modes}

For $\vec{k} \cdot \vec{j}_n \neq 0 \neq \vec{k} \cdot \vec{j}_s$ and $\vec{k} \times \vec{v}_n = 0$, there are two degenerate pairs of solutions to the equations of motion, 
a pair that corresponds to first sound and a pair that corresponds to fourth sound.  With
\begin{align}
f_s \equiv \frac{\rho_s}{\rho},
\end{align}
to first order in $f_s$, first sound frequencies are given by
\begin{align}
\frac{\omega_1^2}{k^2} = c_1^2 + f_s \left[c_1^2 - 2 \widetilde{c}^2 + \frac{\widetilde{c}^4}{c_1^2}  + w_{ll}^{(0)}\frac{\partial \lambda}{\partial \rho}\left(\frac{\widetilde{c}^2}{c_1^2}-1 \right) \right] ,
\end{align}
and fourth sound frequencies are given by
\begin{align}
\frac{\omega_4^2}{k^2} = f_s\left( c_0^2 - \frac{\widetilde{c}^4}{c_1^2}- w_{ll}^{(0)} \frac{\partial \lambda}{\partial \rho} \frac{\widetilde{c}^2}{c_1^2} \right) .
\end{align}
Here, the velocities $c_0$, $c_1$, and $\widetilde{c}$ satisfy
\begin{align}
c_0^2 \equiv& \rho \frac{\partial \mu}{\partial \rho},\label{c0}\\
c_1^2 \equiv& \frac{\partial P}{\partial \rho} - \frac{\partial \lambda}{\partial \rho}+  \frac{1}{\rho} \left(\frac{\partial \lambda}{\partial w} - \frac{\partial P}{\partial w} \right),\label{c1}\\
\widetilde{c}^2 \equiv& c_0^2 - \frac{\partial \lambda}{\partial \rho}.\label{chat}
\end{align}
If $\sigma$ rather than $s$ were held constant, $c_0$ would be the sound velocity in an ordinary (non-super) liquid, and $c_1$ would be the sound velocity in an ordinary solid.\cite{SearsSasALNS10}  Ref.~\onlinecite{SearsSasSSGen} shows that for $P_a \ll K$ we have $c_1^2 \gg \widetilde{c}^2 \gg c_0^2 $ and strain $w_{ll}^{(0)} \ll 1$.  It is also convenient to define the ``fluid-like'' and ``solid-like'' velocities $c_{lL}$ and $c_{lS}$, which satisfy\cite{SearsSasALNS10}
\begin{align}
c_{lL}^2 \equiv \frac{\partial P}{\partial \rho} - \frac{\partial \lambda}{\partial \rho},\qquad c_{lS}^2 \equiv \frac{1}{\rho} \left(\frac{\partial \lambda}{\partial w} - \frac{\partial P}{\partial w} \right), \label{clLclS}
\end{align}
so that
\begin{align}
c_1^2 = c_{lL}^2 + c_{lS}^2.
\end{align}
For an ordinary solid, the derivatives in \eqref{clLclS} are taken at constant $\sigma$ rather than $s$.

\subsubsection{Transverse Elastic Modes} 
For $\vec{k} \cdot \vec{j}_n = 0 = \vec{k} \cdot \vec{j}_s$ and $\vec{k} \times \vec{v}_n \neq 0$, 
there are two degenerate pairs of elastic modes.  They each have a frequency satisfying 
\begin{align}
\omega_t = k \sqrt{\frac{\mu_V}{\rho_n}},
\end{align}
which is larger than the ordinary (non-super) solid transverse frequency by the factor $\sqrt{\rho/\rho_n}$.  Such an effect, to our knowledge, has not been observed.

\subsection{Diffusive Mode}
\label{DiffusiveSubsection}
For the diffusive mode, we keep the dissipative terms in the equations of motion \eqref{s1}-\eqref{vs1}, so that ${\dot{u}'}_i \neq {v_n}_i'$.  With $w'_{jl} = i k_j u_l'$, rewriting \eqref{s1}-\eqref{vs1} in terms of the variables ${v'_n}_i$, ${v'_s}_i$, $\rho'$, $s'$ and ${u'_i}$ gives
\begin{align}
&\omega s' = k_i s {v_n}_i' - i k^2 \frac{\kappa}{T} \left( \frac{\partial
T}{\partial s} s' +  \frac{\partial T}{\partial \rho} \rho' +  \frac{\partial
T}{\partial w_{jl}} i k_j u_l' \right), \label{sdiff1}\\
&\omega {u_i}' = i {v_n}_i' - \beta k_k \left(\frac{\partial \lambda_{ki}}{\partial s}
s' +  \frac{\partial \lambda_{ki}}{\partial \rho} \rho' +   \frac{\partial \lambda_{ki}}{\partial w_{jl}} i k_j u_l' \right),
\label{udiff1}\\
&\omega \rho' = k_i g_i' = k_i \left(\rho_n {v_n}_i' + \rho_s {v_s}_i' \right),
\label{rhodiff1}\\
&\omega g_i' = -k_k \sigma_{ik}' = k_k \left[\left( \frac{\partial P}{\partial s} \delta_{ik} - 
\frac{\partial \lambda_{ik}}{\partial s} \right)s' \right. \notag\\
&\quad \left. +  \left(
\frac{\partial P}{\partial \rho} \delta_{ik} -  \frac{\partial
\lambda_{ik}}{\partial \rho} \right)\rho' + \left( \frac{\partial
P}{\partial w_{jl}} \delta_{ik} -  \frac{\partial \lambda_{ik}}{\partial w_{jl}}
\right)i k_j u_l' \right], \label{gdiff1} \\
&\omega {v_s}_i' =  k_i \mu' = k_i \left( \frac{\partial \mu}{\partial s}s' + \frac{\partial
\mu}{\partial \rho}\rho' + \frac{\partial \mu}{\partial w_{jl}}i k_j u_l' \right).
\label{vsdiff1}
\end{align}
Recall that we have neglected the viscosity as a higher-order effect in $k^2$ as $k \rightarrow 0$. 
We assume that 
\begin{align}
\omega =-iD_{D}k^{2},
\label{omegadiff}
\end{align}
 where the diffusion constant $D_{D}>0$ is to be determined.  

At first sight this system promises to yield a quintic in $\omega$, associated with the longitudinal modes.  However, the assumption that there is a diffusive mode (whose consistency we must verify) permits us to reduce this to a single linear equation.  In some sense a single diffusive mode is expected, because we have already obtained four pairs of propagating modes.  We detail our procedure because it both illuminates the physics and clarifies the mathematics. 

{\bf (1) Method of Solution. }
Since we take the long wavelength limit, we neglect terms that are higher order in $k$.  
In the present analysis we are merely interested in an order of magnitude estimation so we drop subscripts.  When later solving for the frequency and mode structure we use appropriate subscripts.
   
When written in terms of powers of $k$ (using \eqref{omegadiff}), mass and momentum conservation (eqs.~\eqref{rhodiff1} and \eqref{gdiff1}) imply that
\begin{align}
k^2 \rho' \sim& k g', \label{krhog}\\
k^2 g' \sim& k \widetilde{\sigma}'.\label{kgsigma}
\end{align}
Here we use $\widetilde{\sigma}$ to distinguish a stress from $\sigma$, the entropy/mass.  Combination of \eqref{krhog} and \eqref{kgsigma} yields $k^2 \rho' \sim \widetilde{\sigma}'$.  Since expanding $\widetilde{\sigma}'$ in terms of the other variables gives a term proportional to $\rho'$, for small $k$ the term $k^{2}\rho'$ is negligible, so $\widetilde{\sigma}' \rightarrow 0$ as $k \rightarrow 0$.  The diffusive mode therefore is characterized by a negligible stress deviation.  Physically this means that the fluid-like stress deviation nearly cancels the solid-like stress deviation.  When $\widetilde{\sigma}'$ is expanded in terms of the other variables, the condition $\widetilde{\sigma}'\approx 0$ provides a relationship between $s'$, $\rho'$ and $k u'$.

We now turn to the superfluid equation \eqref{vsdiff1}, which gives
\begin{align}
k^2 v_s' \sim k \mu'.
\label{kvskmu}
\end{align}
We now assume that $\mu' \rightarrow 0$ as $k \rightarrow 0$, to be verified below.  When $\mu'$ is expanded in terms of the other variables, the condition $\mu'\approx 0$ provides a second relationship between $s'$, $\rho'$ and $k u'$.  In the remaining equations, for $s'$ and $u'$, we choose to eliminate $\rho'$ and $u'$ in favor of $s'$.

Neither of the equations for $s'$ or $u'$ (eqs.~\eqref{sdiff1}-\eqref{udiff1}) involve $v'_{s}$.  Hence, on eliminating $\rho'$ and $u'$ in favor of $s'$, eqs.~\eqref{sdiff1} and \eqref{udiff1} involve $s'$ and $v'_{n}$, as well as the unknown $\omega$.  This leaves us with two linear equations for two unknowns: the ratio of $v'_{n}$ to $s'$, and $\omega$.  Once these are determined, we use conservation of mass to relate the still-unknown $v'_{s}$ to $v'_{n}$ and $\rho'$, both of which having been found in terms of $s'$.  We find that at low temperatures $\omega \rho'$ can be neglected relative to $k v_n'$, so that \eqref{rhodiff1} gives $0\approx g'=\rho_{n}v'_{n}+\rho_{s}v'_{s}$.  This is not a result of an analysis in powers of $k$ as $k\rightarrow0$, but rather from relations between various thermodynamic quantities.

In what follows, several Maxwell relations from \eqref{s-depsilon} are used:  
\begin{align}
\frac{\partial \mu}{\partial s} = \frac{\partial T}{\partial \rho}, \quad \frac{\partial \lambda_{ik}}{\partial s} = \frac{\partial T}{\partial w_{ik}}, \quad \frac{\partial \lambda_{ik}}{\partial \rho} = \frac{\partial \mu}{\partial w_{ik}}.
\label{MaxwellRels}
\end{align}
Further, Ref.~\onlinecite{LLElasticity} gives, for the elastic stress,
\begin{align}
\lambda_{ik} = \left(\frac{\partial \lambda}{\partial w} - 2 \mu_V \right)\delta_{ik} w_{ll} + \mu_V \left(w_{ik} + w_{ki} \right),
\label{LLlambda}
\end{align}
where $\partial \lambda/\partial w$ is defined in \eqref{dlambdadw}.  
Since, as in \eqref{staticstrain}, the static strain is isotropic (i.e., $ w_{ik}^{(0)} \sim \delta_{ik}$), eq.~\eqref{LLlambda} implies that the static elastic stress also is isotropic (i.e., $\lambda_{ik}^{(0)} \sim \delta_{ik}$).  Thus $(\partial\lambda_{ik}/\partial\rho)_{w_{jl}}$ and $(\partial\lambda_{ik}/\partial s)_{w_{jl}}$ also are isotropic, which permits us to define
\begin{align}
\frac{\partial T}{\partial w_{ik}} = \frac{\partial \lambda_{ik}}{\partial s} \equiv \frac{\partial \lambda}{\partial s} \delta_{ik}, \quad \frac{\partial \mu}{\partial w_{ik}} = \frac{\partial \lambda_{ik}}{\partial \rho} \equiv \frac{\partial \lambda}{\partial \rho} \delta_{ik}.
\label{MaxwellRelsIso}
\end{align}
Note that eq.~\eqref{LLlambda} gives 
\begin{align}
\frac{\partial \lambda_{ik}}{\partial w_{jl}} = \left(\frac{\partial \lambda}{\partial w} - 2 \mu_V \right)\delta_{ik} \delta_{jl} + \mu_V \left(\delta_{ij}\delta_{kl} + \delta_{kj} \delta_{il} \right).
\label{dlambdaik/dwjl}
\end{align}

{\bf (2) Rewriting Stress Equation. }
%
%
For the isotropic case, using \eqref{dPdw}-\eqref{dlambdadrho}, \eqref{clLclS}, \eqref{MaxwellRels} and \eqref{MaxwellRelsIso}-\eqref{dlambdaik/dwjl}, eq.~\eqref{gdiff1} gives, for negligible total stress,
\begin{align}
0 \approx&  \left(\frac{\partial P}{\partial s}  -  \frac{\partial
\lambda}{\partial s} \right) k_i s' +  \left( \frac{\partial P}{\partial \rho} -  \frac{\partial \lambda}{\partial \rho} \right)
k_i \rho'  
\notag\\
& \qquad   - \left( \frac{\partial \lambda}{\partial w} - \mu_V - \frac{\partial P}{\partial w} \right) i k_i k_l u_l' - \mu_V i k^2 u_i' \notag\\
& \approx  -\frac{\partial \widetilde{\sigma}}{\partial s}  k_i s' +  c_{lL}^2 k_i \rho' - (\rho c_{lS}^2-\mu_V) i k_i k_l u_l' - \mu_V i k^2 u_i',
\label{gdiff3}
\end{align}
where we define 
\begin{align}
\frac{\partial \widetilde{\sigma}}{\partial s} \equiv \frac{\partial \lambda}{\partial s}  -  \frac{\partial P}{\partial s} .
\label{dlambdads}
\end{align}
Since each term of \eqref{gdiff3} except the last is proportional to $k_i$, we have that $u_i'$ is proportional to $k_i$.  Thus, $k_i k_l u_l' = k^2 u_i'$, and \eqref{gdiff3} becomes, on taking the dot product with $k_i/k^2$ and dropping the indices on $k_l u_l'$,
\begin{align}
0 \approx  -\frac{\partial \widetilde{\sigma}}{\partial s}  s' +  c_{lL}^2 \rho' - \rho c_{lS}^2 i k u' .
\label{gdiff3.1}
\end{align}

Further, since $u_i' \sim k_i$, substitution of \eqref{MaxwellRelsIso}-\eqref{dlambdaik/dwjl} into \eqref{udiff1} gives ${v_n'}_i \sim k_i$.  Then, since \eqref{vsdiff1} gives ${v_s'}_i \sim k_i$, the diffusive mode is purely longitudinal (${v_s'}_i \sim {v_n'}_i \sim u'_i \sim k_i$), and we therefore drop indices for $v_s'$, $v_n'$, and $u'$ dotted with $k$.  Moreover, for $u_i' \sim k_i$, \eqref{dlambdaik/dwjl} gives 
\begin{align}
\frac{\partial \lambda_{ki}}{\partial w_{jl}} i k_k k_j u_l' = \frac{\partial \lambda}{\partial w} i k^2 u_i'.
\label{dlambdaik/dwjl2}
\end{align}

{\bf (3) Rewriting $\mu'$ Equation. }
Since we assume that $\mu'\approx0$, we neglect the LHS of \eqref{vsdiff1}; this yields
\begin{align}
0 \approx  \frac{\partial \mu}{\partial s} s' + \frac{\partial \mu}{\partial
\rho} \rho' + \frac{\partial \mu}{\partial w_{jl}} i k_j u_l' . \label{vsdiff2}
\end{align}
Substitution from \eqref{MaxwellRels} and \eqref{MaxwellRelsIso} gives
\begin{align}
0 \approx  \frac{\partial T}{\partial \rho} s' + \frac{\partial \mu}{\partial
\rho} \rho' + \frac{\partial \lambda}{\partial \rho} i k u' .
\label{vsdiff3}
\end{align}
  
{\bf (4) Combining stress and $\mu'$ equations.} Solving \eqref{gdiff3.1} and \eqref{vsdiff3} for $\rho'$ and $u'$ gives
\begin{align}
\rho' = \frac{Y_3}{Y_1} \frac{\rho s'}{s}, \qquad -i k u' = \frac{Y_2}{Y_1} \frac{s'}{s},
\label{rhoANDuTOs}
\end{align}
where we introduce three quantities, each with units of velocity to the fourth power:
\begin{align}
Y_1 &\equiv  \frac{\partial \lambda}{\partial \rho } c_{lL}^2 + c_0^2 c_{lS}^2   \label{Y1},\\
Y_2 &\equiv  s \frac{\partial T}{\partial \rho} {c}_{lL}^2 +  c_0^2 \frac{s}{\rho} \frac{\partial \widetilde{\sigma}}{\partial s} \label{Y2},\\
Y_3 &\equiv  \frac{s}{\rho} \frac{\partial \lambda}{\partial \rho}\frac{\partial \widetilde{\sigma}}{\partial s} - s \frac{\partial T}{\partial \rho} c_{lS}^2       \label{Y3}.
\end{align}
Here we employ \eqref{c0}.  Eq.~\eqref{rhoANDuTOs} holds for any $\omega \sim k^2$.   

Appendix~\ref{PaTOKappendix} uses the results of Ref.~\onlinecite{SearsSasALNS10} to estimate the sizes of $Y_1$, $Y_2$, and $Y_3$.  
With $\theta_D$ the Debye temperature, $k_B$ Boltzmann's constant, and $m_4$ the atomic mass of $^4$He, 
we find 
\begin{align}
Y_1 \approx& -\frac{2 P_a^2}{\rho^2} , \qquad Y_2 \approx - \frac{24 \pi^4}{9} \frac{T^3}{\theta_D^3}\frac{k_B T}{m_4} \frac{P_a}{\rho} , \notag\\
&\quad Y_3 \approx - \frac{24 \pi^4}{9} \frac{T^3}{\theta_D^3}\frac{k_B T}{m_4} \frac{K}{\rho} . \label{Y1Y2Y3}
\end{align}
Note that $Y_1$ is independent of $T$.  To evaluate these we take 
$\theta_D \approx 25$~K,\cite{CrepHeyLee} $m_4 \approx 6.7 \times 10^{-27}$~kg, $\rho \approx 2 \times 10^3$~kg/m$^3$, $P_a \approx 30$~bar, and $K \approx 300$~bar.\cite{SearsSasSSGen}  Further, following evidence that  a supersolid phase of $^4$He can only exist at $T<55$~mK,\cite{ShevDayBeam10} we take $T \approx 50$~mK.  Then, eq.~\eqref{Y1Y2Y3} yields
\begin{align}
Y_1 \approx& -4.5 \times 10^6 \,\,\frac{{\rm m}^4}{{\rm s}^4} , \quad Y_2 \approx - 3.25 \times 10^{-1} \,\,\frac{{\rm m}^4}{{\rm s}^4},\notag\\
 &\quad \qquad Y_3 \approx - 3.25  \,\,\frac{{\rm m}^4}{{\rm s}^4}, 
 \label{evalYs}
\end{align}
so that $Y_1 \gg Y_3 \gg Y_2$.  This inequality applies for any $T<55$~mK, and therefore applies at any temperature relevant to supersolid $^4$He experiments subject to $P_a \ll K$.

{\bf (5) Rewriting $s'$ and $u_{i}'$ Equations. }
Substituting \eqref{rhoANDuTOs}, \eqref{dlambdaik/dwjl2}, and \eqref{MaxwellRelsIso} into \eqref{sdiff1}, and into \eqref{udiff1} multiplied by $-i k s Y_1/Y_2$, yields
\begin{align}
 &\omega s' = k s {v_n'} - i k^2 \frac{\kappa}{T} \left( \frac{\partial T}{\partial s}   +  \frac{\rho Y_3}{s Y_1}  \frac{\partial T}{\partial \rho} - \frac{Y_2}{s Y_1} \frac{\partial \lambda}{\partial s}   \right) s', \label{sdiff2}\\
&\omega  s' = k \frac{s Y_1}{Y_2} {v_n'} + i  k^2 \beta \left( \frac{s Y_1}{Y_2} \frac{\partial \lambda}{\partial s}
  + \frac{\rho Y_3}{Y_2} \frac{\partial \lambda}{\partial \rho} -   \frac{\partial \lambda}{\partial w} \right) s'.
\label{udiff2}
\end{align}
We simplify \eqref{sdiff2}-\eqref{udiff2} by the following argument.  We take $\partial \lambda/\partial s$ to have the same linear $T$-dependence as $\partial P/\partial s$ in a harmonic solid, or $\partial \lambda/\partial s \sim T$.  Then, since $s \sim T^3$ and $Y_2,Y_3 \sim T^4$, all terms $\sim \kappa$ in \eqref{sdiff2} have the same temperature dependence, and the same is true for all terms $\sim \beta$ in \eqref{udiff2}.  Thus, since 
$Y_1 \sim 10^6 \times Y_3$ and $Y_3 \sim 10 \times Y_2$, in the parentheses of \eqref{sdiff2} and \eqref{udiff2} the first term dominates.  Thus \eqref{sdiff2}-\eqref{udiff2} approximately give, on rearranging,
\begin{align}
 &\left(\omega +i k^2 \frac{\kappa}{T} \frac{\partial T}{\partial s} \right)s' = k s {v_n'}, \label{sdiff2.1}\\
&\left(\omega - i  k^2 \beta  \frac{s Y_1}{Y_2} \frac{\partial \lambda}{\partial s} \right)s' = k \frac{s Y_1}{Y_2} {v_n'} .
\label{udiff2.1}
\end{align}


Subtracting \eqref{udiff2.1} from \eqref{sdiff2.1} and dividing by $ks$ yields
\begin{align}
 {v_n}'  =& i k \frac{s'}{s} \left(\frac{1}{T} \frac{\partial T}{\partial s} \right) \left(\kappa + \frac{Y_1}{Y_2} \beta s T \frac{\partial \lambda}{\partial T}   \right) ,
 \label{vnTOsdiff}
 \end{align}
which holds for any $\omega\sim k^{2}$.  Here we use $(\partial \lambda/\partial s)/(\partial T/\partial s) = \partial \lambda/\partial T$, where $\rho$ and $w_{ik}$ are implicitly held constant for each derivative.  Hence, eqs.~\eqref{rhoANDuTOs} and \eqref{vnTOsdiff} show the ratios of ($\rho'$, $k u'$, $v_n'$) to $s'$ to be frequency independent.  

We now find the frequency of the diffusive mode using \eqref{sdiff2.1} and \eqref{udiff2.1}.  Mass conservation from \eqref{rhodiff1} then relates $v_s'$ and $s'$, thus yielding all variables in terms of $s'$. 

\subsubsection{Diffusive Mode Frequency}
\label{DiffFreqSubsubsection}
Cross-multiplication of \eqref{sdiff2.1} and \eqref{udiff2.1} yields 
\begin{align}
 & \omega   + i k^2 \frac{\kappa}{T} \frac{\partial T}{\partial s}  =  \frac{ Y_2}{Y_1} \omega - i  k^2 \beta s  \frac{\partial \lambda}{\partial s} .
\label{DiffusiveFreq1}
\end{align}
The frequency of the diffusive mode thus is 
\begin{equation}
 \omega = -i k^2 \frac{\bar{\kappa}}{T}\frac{\partial T}{\partial s},
\label{DiffFreqApprox}
\end{equation}
where
\begin{align}
 \bar{\kappa} \equiv& \left[ \frac{\displaystyle \kappa + \beta s T\frac{\partial \lambda/\partial s}{\partial T/\partial s} } {\displaystyle 1 - \frac{Y_2}{Y_1}}\right] 
 \approx \kappa + \beta s T \frac{\partial \lambda}{\partial T} .
\label{kappabar}
\end{align}
Here we use $Y_1 \gg Y_2$.  Recall that $\partial \lambda/\partial T$ is taken at constant $\rho$ and $w_{ik}$.  The frequency thus has a part associated with thermal diffusion ($\sim \kappa$) and a part associated with lattice diffusion ($\sim \beta$).

Before finding the full mode structure, it is worth commenting on \eqref{DiffFreqApprox}-\eqref{kappabar}.  If $\beta s T (\partial \lambda/\partial T) \ll \kappa$, then we have $D_D \rightarrow (\kappa/T)(\partial T/\partial s)$, as for ordinary thermal diffusion.  As noted above, however, $\mu$ is constant in the long wavelength limit for the diffusive mode of the supersolid (to be verified below), unlike in the case of a fluid or ordinary solid.  Thus, even if the frequency were precisely as for normal thermal diffusion, the mode structure (e.g., $v_n'/v_s'$, etc.) 
would nonetheless be different than for the usual case.

\subsubsection{$v_n'$, $v_s'$, and $\dot{u}'$ in the Diffusive Mode}
\label{vnvsudotSection}

Eq.~\eqref{rhodiff1} gives
\begin{align}
v_s' = \frac{\omega \rho'}{k \rho_s}  - \frac{\rho_n v_n'}{\rho_s}.
\label{vs0}
\end{align}  

By \eqref{rhoANDuTOs}, the first term on the RHS of \eqref{vs0} is given by 
\begin{align}
\frac{\omega \rho'}{k \rho_s} =& \frac{\omega \rho s'}{k \rho_s s}  \left[\frac{Y_3}{Y_1}\right] .
\label{vsrho}
\end{align}
Further, using \eqref{kappabar} and $Y_1 \gg Y_2$, eq.~\eqref{vnTOsdiff} can be written as 
\begin{align}
 {v_n}'  =& i k \frac{\bar{\kappa}}{T}  \frac{\partial T}{\partial s} \frac{s'}{s} \left[1 + \left( \frac{Y_1}{Y_2} - 1 \right) \frac{\beta s T}{\bar{\kappa}} \frac{\partial \lambda}{\partial T}     \right]  \notag\\
  \approx& - \frac{\omega s'}{k s}   \left[1+  \frac{Y_1}{Y_2}  \frac{\beta s T}{\bar{\kappa}}  \frac{\partial \lambda}{\partial T}      \right].
\label{vnTOsapprox}
\end{align}
Thus the second term on the RHS of \eqref{vs0} is given by 
\begin{align}
- \frac{\rho_n v_n'}{\rho_s} \approx& \frac{\omega \rho_n s'}{k\rho_s s}   \left[ 1+  \frac{Y_1}{Y_2}  \frac{\beta s T}{\bar{\kappa}}  \frac{\partial \lambda}{\partial T}   \right] .
\label{vsvn}
\end{align}
Since experiments\cite{RitRep09} indicate that $\rho_n \gtrsim 0.8 \rho$, on using $Y_1 \gg Y_3$, eqs.~\eqref{vsrho} and \eqref{vsvn} give $-({\rho_n}/{\rho_s}) v_n' \gg (\omega \rho'/k \rho_s) $.  Eq.~\eqref{vs0} therefore becomes, on employing \eqref{vnTOsdiff},
\begin{align}
v_s' \approx - \frac{\rho_n}{\rho_s} v_n' = -  i k \frac{\rho_n s'}{\rho_s s} \left(\frac{1}{T} \frac{\partial T}{\partial s} \right) \left(\kappa + \frac{Y_1}{Y_2} \beta s T \frac{\partial \lambda}{\partial T}   \right) ,
\label{vsTOvn}
\end{align} 
or, equivalently, $g' = \rho_n v_n' + \rho_s v_s' \approx 0$.  Thus the superfluid velocity is opposite the normal velocity, with a weighting given by $\rho_n/\rho_s$. 
Since $\rho_n/\rho_s \geq 4$ we approximately have $|v_s'| \gg |v_n'|$.  Note that \eqref{vsTOvn} explicitly relates $v'_{s}$ to $s'$, thus completely specifying the eigenmode. 


We now verify the assumption that $\mu' \rightarrow 0$ for $k \rightarrow 0$.  We do so by showing the LHS of \eqref{vsdiff1} to be negligible compared to any given term on the RHS (e.g., $k^2 v_s' \ll (\partial \mu/\partial s) s'$).  Counting powers of $k$, eqs.~\eqref{vsTOvn} and \eqref{vnTOsdiff} give $v_s' \sim v_n' \sim k s'$.  Thus $k^2 v_s' \sim k^3 s' \ll (\partial \mu/\partial s) s'$, which shows the consistency of the assumption.

Furthermore, using eq.~\eqref{rhoANDuTOs} to write the lattice velocity $\dot{u}'$ gives
\begin{align}
\dot{u}' = - i \omega u' = \frac{Y_2}{Y_1} \frac{\omega s'}{k s}.
\end{align}
Comparison to \eqref{vnTOsapprox} yields
\begin{align}
v_n' = - \dot{u}'  \frac{Y_1}{Y_2}  \left[1+  \frac{Y_1}{Y_2}  \frac{\beta s T}{\bar{\kappa}} \left. \frac{\partial \lambda}{\partial T}\right|_{\rho, w_{ik}}      \right].
\label{vnTOuDIFF}
\end{align}
Since $Y_1 \gg Y_2$, unless $[\beta s T  (\partial \lambda/\partial T)/\bar{\kappa}]  \approx -Y_2/Y_1$ (an unlikely coincidence), we have $v_n' \approx - \dot{u}'  ({Y_1}/{Y_2})  \gg   |\dot{u}'|$.  Then, by \eqref{vsTOvn}, for $\rho_n \gg \rho_s$, we have $|v_s'| \gg |v_n'| \gg |\dot{u}'|$.  Since $v_n' \neq \dot{u}'$, mass motion is distinct from lattice motion.  

\section{Summary}
\label{Summary}
We have re-examined the supersolid hydrodynamics of Andreev and Lifshitz, including the effects of nonzero applied pressure $P_{a}$.   For $P_a \neq 0$, a solid responds with both lattice stress $\lambda_{ik}$ and internal pressure $P$.  The dependence of $P$ and $\lambda_{ik}$ on $P_a$ is found in Ref.~\onlinecite{SearsSasALNS10}, and employed here to describe the eigenmodes.  We first summarized the results for the four degenerate pairs of longitudinal and transverse elastic mode frequencies (including fourth sound); because we include $P_{a}$ and the associated strain, the results differ somewhat from those of previous work.  In addition, again including $P_{a}$ and the associated strain, in the long-wavelength limit we have obtained the previously-unstudied diffusive eigenmode.  
 
The diffusive mode frequency, under certain conditions, is similar to the frequency of ordinary thermal diffusion.  However, the mode involves no deviations in net stress or net chemical potential, so its properties differ from ordinary thermal diffusion.  To produce zero net stress deviation, the solid-like elasticity component is cancelled by the previously neglected fluid-like component associated with lattice defects.  To produce zero net chemical potential deviation, the temperature and pressure deviations must be related. With zero net stress deviation we find that at low temperature there also is zero net momentum.  With the normal fluid density dominating the superfluid density, this means that the superfluid velocity is much larger than the normal fluid velocity.  Because the lattice displacement is coupled to the elastic strain with a large coefficient, but the normal fluid velocity is coupled to the fluid-like strain (a pressure) with a small coefficient, zero net stress deviation implies that the normal fluid velocity is much greater than the lattice velocity.  This is an unusual phenomenon, since in the other modes the lattice velocity and normal fluid velocity are nearly equal.  

A previous work studied the lattice diffusion mode for a normal solid having distinct velocities associated with momentum (e.g., the normal fluid velocity) and lattice elasticity.\cite{SearsSasALNS10}  The motivation was to consider that the time-delay in the pressurization experiments of Ref.~\onlinecite{RitRep09} might be due to that mode, under the assumption that the sample studied is not supersolid.  Similar considerations can be made for the diffusive mode we have just studied, because both modes are diffusive in nature, and thus would show a dependence on the sample thickness $d$ as $d^{2}$.  A study of this dependence would be of interest, to confirm that the effect observed in Ref.~\onlinecite{RitRep09} is diffusive in nature.  



\appendix
\section{Relative Sizes of $Y_1$, $Y_2$ and $Y_3$ for Small $P_a$}
\label{PaTOKappendix}
Using the results of Refs.~\onlinecite{SearsSasALNS10} and \onlinecite{SearsSasSSGen} and estimating certain thermodynamic derivatives under the condition $P_a \ll K$ allows us to estimate the sizes of $Y_1$, $Y_2$, and $Y_3$.  
Ref.~\onlinecite{SearsSasALNS10} gives for a normal solid, to lowest order in $P_a/K$,
\begin{align}
&w_{ll}^{(0)} = - \frac{P_a}{K},\label{wllAppend}\\
&\frac{\partial \lambda}{\partial \rho} = - c_{lL}^2 =  \frac{V P_a}{\rho K} \left. \frac{\partial K}{\partial V}\right|_{\sigma, w_{ik}, N} ,\label{dlambdadrhoAppend}\\
&\frac{\partial P}{\partial \rho} = \frac{V^2 P_a^2}{2 \rho K^2} \left. \frac{\partial ^2 K}{\partial V^2} \right|_{\sigma, w_{ik}, N},\label{dPdrhoAppend}\\
&\frac{\partial P}{\partial w} =  -P_a \left(1 -  \frac{V}{K} \left. \frac{\partial K}{\partial V}\right|_{\sigma, w_{ik}, N} \right),\label{dPdwAppend}\\
&c_1^2 = c_{lS}^2 = \frac{\partial \lambda}{\partial w} = \frac{K+\frac{4}{3} \mu_V}{\rho}. \label{c1Append}
\end{align}
Although Ref.~\onlinecite{SearsSasALNS10} evaluates the derivatives of $\lambda_{ik}$ and $P$ at constant $\sigma$ rather than $s$, at $T = 0$ we have $\sigma \approx 0 \approx s$, so holding either quantity constant should give nearly equivalent results for supersolid $^4$He.  All derivatives of $K$ here are taken at constant $\sigma$, $w_{ik}$, and $N$, so we now drop the subscripts.  Ref.~\onlinecite{SearsSasSSGen} also finds
\begin{align}
c_0^2 \approx \frac{V P_a^2}{\rho K^2}\left(\frac{V}{2}  \frac{\partial ^2 K}{\partial V^2}  +   \frac{\partial K}{\partial V} \right).
\label{c0Append}
\end{align}

Constant $w_{ik}$ constant is equivalent to constant density of lattice sites.  Because $K$ is a measure of the material stiffness, one expects $K$ to increase as $V$ decreases, for constant $N$ and $w_{ik}$, i.e., $\partial K/\partial V<0$.  Then by \eqref{dlambdadrhoAppend} and \eqref{dPdwAppend} we have $\partial \lambda/\partial \rho <0$ and $\partial P/\partial w<0$.  

For the putative supersolid, we approximate $\partial \widetilde{\sigma}/\partial s$ using \eqref{s-GibbsDuhem}, \eqref{MaxwellRels}, and \eqref{MaxwellRelsIso}:
\begin{align}
\frac{\partial \widetilde{\sigma}}{\partial s} =& \frac{\partial \lambda}{\partial s} - \frac{\partial P}{\partial s}
\approx \frac{\partial \lambda}{\partial s} - s \frac{\partial T}{\partial s} - \rho \frac{\partial T}{\partial \rho} - w_{ll}^{(0)} \frac{\partial \lambda}{\partial s}.
\end{align}
Recall that, unless otherwise specified, derivatives with respect to $\rho$, $s$ or $w_{ik}$ are taken with the other two held constant. 
Eq.~\eqref{wllAppend} shows that for $P_a \ll K$, we have $w_{ll}^{(0)} \ll 1$.  
Also,\cite{SearsSasALNS10} as noted earlier, $\rho (\partial T/\partial \rho) \approx \gamma s (\partial T/\partial s)$, where $\gamma \approx 10$.  Thus,
\begin{align}
\frac{\partial \widetilde{\sigma}}{\partial s} \approx \frac{\partial \lambda}{\partial s} - (1 + \gamma^{-1}) \rho \frac{\partial T}{\partial \rho}.
\label{dsigmadsAppend}
\end{align}
On neglecting $\mu_V$, eqs.~\eqref{LLlambda} and \eqref{wllAppend} give $\partial \lambda/\partial s \approx (\partial K/\partial s) w_{ll}^{(0)} \approx (P_a/K)(\partial K/\partial s)$.

Substitution of \eqref{dlambdadrhoAppend}, \eqref{c1Append}, \eqref{c0Append} and \eqref{dsigmadsAppend} into \eqref{Y1}-\eqref{Y3} gives, to lowest order in $P_a/K$, 
\begin{align}
Y_1 =& \frac{\partial \lambda}{\partial \rho } c_{lL}^2 + c_0^2 c_{lS}^2  \notag\\
\approx& -\frac{V^2 P_a^2}{\rho^2 K^2}\left[ \left( \frac{\partial K}{\partial V}\right)^2 \right.  \notag\\
&\qquad  \left.+ \left({K+\frac{4}{3} \mu_V}\right)\left(\frac{1}{2}  \frac{\partial ^2 K}{\partial V^2}  +   \frac{1}{V}\frac{\partial K}{\partial V} \right)\right]\label{Y1Append},\\
Y_2 =&  s \frac{\partial T}{\partial \rho} {c}_{lL}^2 +  c_0^2 \frac{s}{\rho} \frac{\partial \widetilde{\sigma}}{\partial s} \approx - s \frac{\partial T}{\partial \rho} \frac{V P_a}{\rho K}  \frac{\partial K}{\partial V}  \label{Y2Append},\\
Y_3 =&  \frac{s}{\rho} \frac{\partial \lambda}{\partial \rho}\frac{\partial \widetilde{\sigma}}{\partial s} - s \frac{\partial T}{\partial \rho} c_{lS}^2 \approx - s \frac{\partial T}{\partial \rho} \frac{K+\frac{4}{3} \mu_V}{\rho} \label{Y3Append}.
\end{align}
Note that all terms $\sim \partial \widetilde{\sigma}/\partial s$ are higher order in $P_a/K$ and therefore are neglected. Approximating $K$ to be linear in $V$ and neglecting $\mu_V$, eqs.~\eqref{Y1Append}-\eqref{Y3Append} give 
\begin{align}
&Y_1 \approx -\frac{2 P_a^2 }{\rho^2}, \quad
 Y_2 \approx - s \frac{\partial T}{\partial \rho} \frac{P_a}{\rho}  , \quad Y_3 \approx - s \frac{\partial T}{\partial \rho} \frac{K}{\rho} .
 \label{Y1-Y3}
\end{align}
For $P_a \ll K$ we have $Y_3 \gg Y_2$.  

To approximate the relative magnitudes of $Y_3$ and $Y_2$ to $Y_1$, 
we now find an explicit form for $s (\partial T/\partial \rho)_s$.

At low temperatures phonon gas statistical mechanics gives
\begin{align}
s = \frac{2 \pi^2 k_B^4 T^3}{15 \hbar^3 \bar{u}^3},
\label{sLL}
\end{align}
where $\bar{u}$ is an average sound velocity and $k_B$ is the Boltzmann constant.  Further,\cite{SearsSasSSGen}
\begin{align}
\frac{\partial T}{\partial \rho} \approx \frac{ T}{ \bar{u}} \frac{\partial \bar{u}}{\partial \rho} \approx \frac{10 T}{3 \rho}.
\label{dTdrhoAppend}
\end{align}
Combining \eqref{sLL} and \eqref{dTdrhoAppend} gives
\begin{align}
s\frac{\partial T}{\partial \rho} \approx  \frac{4 \pi^2 k_B^4 T^4}{9 \rho \hbar^3 \bar{u}^3} .
\label{sdTdrhoAppend}
\end{align}
In terms of the Debye temperature $\theta_D \approx (6 \pi^2 n_{\nu})^{1/3} (\hbar \bar{u}/k_B)$, where $n_{\nu}$ is the number density of vibrations (essentially one per lattice site),
\begin{align}
s\frac{\partial T}{\partial \rho} \approx \frac{24 \pi^4 k_B T^4}{9 (\rho/n_{\nu}) \theta_D^3} \approx  \frac{24 \pi^4}{9}\frac{ T^3}{\theta_D^3} \frac{k_B T}{m_4} .
\label{sdTdrhoAppend2}
\end{align}
Here $m_4$ is the atomic mass of $^4$He, and we have taken $m_4 n_{\nu} \approx \rho$.  Eq.~\eqref{sdTdrhoAppend2} substituted into \eqref{Y1-Y3} gives \eqref{Y1Y2Y3}.


\begin{thebibliography}{}

\bibitem{AL69} A. F. Andreev and I. M. Lifshitz, Sov. Phys. JETP 29, 1107 (1969).
\bibitem{Thouless69} D. J. Thouless, Ann. Phys. (N.Y.) 52, 403 (1969).  This contains the remark that, for a lattice of bosons, vacancies could be ``in the lowest Bloch state with a finite probability, so the system would be `super' but not `fluid'\,''. 
\bibitem{Chester70} G.V. Chester, Phys. Rev. A 2, 256 (1970).
\bibitem{Leggett70} A. J. Leggett, Phys. Rev. Lett. 25, 1543 (1970).  
\bibitem{BalCau08} For a recent review, see S. Balibar and F. Caupin, J. Phys. Cond. Mat. 20, 173201(2008).

\bibitem{KC1} E. Kim and M. Chan, Nature (London) 427, 225 (2004).  
\bibitem{KC2} E. Kim and M. Chan, Science 305, 1941 (2004).  
\bibitem{RR1} A. S. C. Rittner and J. D. Reppy, Phys. Rev. Lett. 97, 165301 (2006).  
\bibitem{Shira1} M. Kondo, S. Takada, Y. Shibayama, and K. Shirahama, J. Low Temp. Phys. 148, 695 (2007).  
\bibitem{Koj1} Y. Aoki, J. C. Graves, and H. Kojima, Phys. Rev. Lett. 99, 015301 (2007).  
\bibitem{Penzev} A. Penzev, Y. Yasuta, and M. Kubota, J. Low Temp. Phys. 148, 677 (2007).  
\bibitem{RR2} A. S. C. Rittner and J. D. Reppy, Phys. Rev. Lett. 98, 175302 (2007).  
\bibitem{RR3} A. S. C. Rittner and J. D. Reppy, Phys. Rev. Lett. 101, 155301(2008).  
\bibitem{Lin07} X. Lin, A. C. Clark, M. H. W. Chan, Nature 449, 1025 (2007).  
\bibitem{ClarkWestChan07} A. C. Clark, J. T. West, and M. H. W. Chan, Phys. Rev. Lett.  99, 135302 (2007).
\bibitem{DayBeamishNature07} J. Day and J. Beamish, Nature 450, 853 (2007).  
\bibitem{ChanScience08} M. H. W. Chan, Science 319, 1207 (2008).
\bibitem{Day06} James Day and John Beamish, Phys. Rev. Lett. 96, 105304 (2006),
\bibitem{ShevDayBeam10} O. Syshchenko, J. Day, and J. Beamish, Phys. Rev. Lett. 104, 195301 (2010).
\bibitem{Hunt09} B. Hunt {\it et al}, Science 324, 632 (2009).
\bibitem{Choi10} H. Choi {\it et al}, N. Phys. 6, 424 (2010). 
\bibitem{RitRep09} S. C. Rittner and J. D. Reppy, J. Phys. Conf. Ser. 150, 032089 (2009).
\bibitem{SearsSasALNS10} M. R. Sears and W. M. Saslow, ``Andreev-Lifshitz Hydrodynamics Applied to an Ordinary Solid under Pressure,'' unpublished. 

\bibitem{SasJolad} W. M. Saslow and S. Jolad, Phys. Rev. B 73, 092505 (2006).
\bibitem{GalliSas} D. E. Galli, L. Reatto and W. M. Saslow, Phys. Rev. B 76, 052503 (2007).
\bibitem{Saslow77} W. M. Saslow, Phys. Rev. B 15, 173 (1977).
\bibitem{Liu} M. Liu, Phys. Rev. B. 18, 1165 (1978).  
\bibitem{YooDorsey} C.-D. Yoo and A. T. Dorsey, Phys. Rev. B 81, 134518 (2010).
\bibitem{SearsSasSSGen} M. R. Sears and W. M. Saslow, ``Generation Efficiencies for Propagating Modes in a Supersolid,'' unpublished. 
\bibitem{SaslowNote} Ref.~\onlinecite{Saslow77} finds equations of motion identical to those of Ref.~\onlinecite{AL69}, but also includes nonlinear terms.  However, Ref.~\onlinecite{Liu} notes that Ref.~\onlinecite{Saslow77} employs the non-Galilean $\vec{j}_{s}=\rho_{s}\vec{v}_{s}$ in place of the Galilean $\vec{j}_{s}=\rho_{s}(\vec{v}_{s}-\vec{v}_{n})$.  This does not affect the equations of motion until the normal modes are calculated.



\bibitem{Atkins59} K. R. Atkins, Phys. Rev. 113, 962 (1959).
\bibitem{RudnickShapiro62} I. Rudnick and K. A. Shapiro, Phys. Rev. Lett. 9, 191
(1962).
\bibitem{ViscosityFootnote} The deviation of the momentum flux tensor $\Pi_{ik}'$ (eq.~\eqref{s-Pi}) has a term from the viscosity $\sim \eta_{iklm} k_m v_l'$ and a term from the stress tensor $\sim \lambda'_{ki}$.  Ref.~\onlinecite{LLElasticity} gives $\lambda'_{ik} \sim i k_j u_l'$.  For both the propagating modes and the diffusive mode, we find that $u_l' \sim {v'_n}_l/\omega$. 
Thus, for both the propagating modes ($\omega \sim k$) and the diffusive mode ($\omega \sim k^2$) the term in $\Pi_{ik}'$ due to viscosity is, at the least, of order $k$ relative to the term $\lambda_{ki}'$, and is therefore neglected in the long wavelength limit.
\bibitem{LLElasticity} L. D. Landau and E. M. Lifshitz, {\it Theory of
Elasticity}, 3rd ed., Pergamon, Oxford (1986).
\bibitem{CrepHeyLee} R. H. Crepeau, O. Heybey, D. M. Lee, and S. A. Strauss, Phys. Rev. A 3, 1162 (1971).

\end{thebibliography}
\end{document}